\documentclass[aps,prl,showpacs,twocolumn]{revtex4-1}
\usepackage{amssymb}
\usepackage{graphicx}
\usepackage{colordvi}
\usepackage{color}
\usepackage{dcolumn,amsmath,amsthm,amscd,amsfonts,amssymb,epsfig,graphics,graphicx,eucal}
\usepackage{amsmath}
\usepackage{amssymb}
\usepackage{latexsym}
\usepackage{graphicx}
\usepackage{epsfig}
\usepackage{epstopdf}
\usepackage{upgreek} 	%upright greek letters
\begin{document}
\title{The effects of spin-dependent interactions on polarisation of bright polariton solitons}

\author{M.~Sich}
\affiliation{Department of Physics and Astronomy, University of Sheffield,
Sheffield S3 7RH, United Kingdom}
\author{F.~Fras}
\affiliation{Department of Physics and Astronomy, University of Sheffield,
Sheffield S3 7RH, United Kingdom}

\author{J.~K.~Chana}
\affiliation{Department of Physics and Astronomy, University of Sheffield,
Sheffield S3 7RH, United Kingdom}

\author{A.~V.~Gorbach}
\affiliation{Department of Physics, University of Bath, Bath, BA2 7AY, United Kingdom}
\author{S.~V.~Gavrilov}
\affiliation{Institute of Solid State Physics RAS, Chernogolovka,Russia,142432}
\author{R.~Hartley}
\affiliation{Department of Physics, University of Bath, Bath, BA2 7AY, United Kingdom}
\author{E.~A.~Cerda-M\'endez}
\affiliation{Paul-Drude-Institut f\"ur Festk\"orperelektronik, Berlin, Germany}
\author{K.~Biermann}
	\affiliation{Paul-Drude-Institut f\"ur Festk\"orperelektronik, Berlin, Germany}
\author{R.~Hey}
	\affiliation{Paul-Drude-Institut f\"ur Festk\"orperelektronik, Berlin, Germany}
\author{P.~V.~Santos}
	\affiliation{Paul-Drude-Institut f\"ur Festk\"orperelektronik, Berlin, Germany}
\author{M.~S.~Skolnick}
\affiliation{Department of Physics and Astronomy, University of Sheffield, Sheffield S3 7RH, United Kingdom}
\author{D.~V.~Skryabin}
\affiliation{Department of Physics, University of Bath, Bath, BA2 7AY, United Kingdom}
\author{D.~N.~Krizhanovskii}
\affiliation{Department of Physics and Astronomy, University of Sheffield, Sheffield S3 7RH, United Kingdom}

\begin{abstract}
We report on the spin properties of bright polariton solitons supported by an external pump to compensate losses.
We observe robust circularly polarised solitons when a circularly polarised pump is applied, a result attributed to phase synchronisation between nondegenerate TE and TM polarised polariton modes at high momenta. For the case of a linearly polarised pump either $\sigma^+$ or $\sigma^-$ circularly polarised bright solitons can be switched on in a controlled way by a $\sigma^+$ or $\sigma^-$ writing beam respectively.  This feature arises directly from the widely differing interaction strengths between co- and cross-circularly polarised polaritons. In the case of orthogonally linearly polarised pump and writing beams, the soliton emission on average is found to be unpolarised, suggesting strong spatial evolution of the soliton polarisation, a conclusion supported by polarisation correlation measurements. The observed results are in agreement with theory, which predicts stable circularly polarised solitons and unstable linearly polarised solitons resulting in spatial evolution of their polarisation.

\end{abstract}

\maketitle

\emph{Introduction.---} Solitary waves (solitons) are well known and broadly investigated across many areas of modern physics. They have attracted particular recent attention  in  nonlinear optics \cite{agr,amp} and cold atom physics 
\cite{becsol1,becsol2}. Optical and matter wave solitons are formed when photon or particle interactions are able to compensate for the dispersive spreading of wave packets. Semiconductor microcavities, where strong exciton-photon coupling results in the formation of half-light half-matter quasiparticles (2D-polaritons), have attracted much attention recently \cite{book, book2}. Polaritons have been used to predict and demonstrate parametric instability \cite{krizh2000,gip}, bistability \cite{tred,baas},  condensation \cite{kasprzak,balili},  superfluidity \cite{AmoNP2009}, dark \cite{yulin,amo, *Adrados2011} and bright \cite{prl11,pla,sich,prl11,blochncomm} polariton solitons.

Nonlinear spin-dependent interactions of co- and cross-polarised photons may have different strengths \cite{boyd}. Interactions of co- and cross-polarised polaritons not only differ in strength, but also have opposite signs such that polaritons with parallel spins repel and polaritons with opposite spins attract, though the attraction is about one order of magnitude weaker \cite{book,book2}. Importantly,  polariton-polariton interactions lead to nonlinearities which are 2-3 orders of magnitude larger than the non-polaritonic ones achieved in the weak coupling regime. Furthermore, the bright polariton solitons reported recently \cite{sich} can be switched on and off on a picosecond timescale and have well defined size (2-4~$\upmu$m) and amplitude, determined by the interactions and cavity parameters.  As a result, they may create opportunities in ultrafast all-optical digital signal processing based on soliton phenomena \cite{barland,pedaci},  where spin-based architectures have considerable potential \cite{Sarma}. 
 The spin properties of polariton solitons we investigate here are further relevant for the development of polariton logic elements integrated on a micrometer lengthscale and operating at relatively low pump powers \cite{LiewNeurons}. 
  Importantly,the bright polariton solitons  propagate in a dissipative environment where losses are fully compensated by gain from an external continuous wave (CW) pump, permitting scalability.
  % of such polariton circuits.

Interplay of the two polariton spin components naturally leads to a variety of possible nonlinear states in the polariton system, expanding the possibilities to control its dynamics and response to external stimuli. In particular,  spin multistability \cite{gipmult} of exciton-polariton states has recently been   reported\cite{Sarkar2010,Paraiso}, expanding on previous research into nonlinear polarisation effects in optical resonators \cite{gipmult,kitano,cecchi}. Polariton spin switching  has also been demonstrated \cite{shel,AmoNPhot2010,Adrados2011, Gavrilov2013a}. More recently, conservative (i.e. without gain) dark polariton (spinor) half-solitons were reported \cite{darkspin}.

We find that when the background CW pump and the local pulsed writing beam (WB) which triggers the soliton are circularly co-polarised, then solitons with the same circular polarisation are readily excited. Once triggered, the soliton maintains its polarisation during propagation over macroscopic distances, indicating quenching of TE-TM polariton mode splitting at high momenta as a result of phase synchronisation between the polariton modes.  Next, we demonstrate that if the pump is linearly polarised, then  solitons with either $\sigma^+$ or $\sigma^-$ circular polarisation can be triggered by a WB with the corresponding circular polarisation due to the very different interaction strengths between co and cross-circularly polarised polaritons.
% The latter arrangement creates an opportunity for all-optical signal processing using solitons in the cross-circularly polarised states as data bits. 
Theoretically we find that linearly polarised solitons are unstable. As a result, when both the pump and WB are linearly polarised, either the soliton evolves into a circularly polarised state or its polarisation rapidly oscillates in space.

The details of the experimental arrangements required for soliton observation in the GaAs-based microcavity sample can be found elsewhere \cite{sich}.
 
\emph{Numerical model.---} Neglecting excitonic kinetic energy and the spin-orbit interaction, we obtain the well known equations for the amplitudes of excitonic oscillators with positive and negative spins \cite{sich}:

\begin{eqnarray}
&& \partial_t\psi_{\pm} + \left(\gamma_{e}- i\delta_e + i|\psi_{\pm}|^{2} - iV
|\psi_{\mp}|^{2} \right) \psi_{\pm}\\ &&  \nonumber = i \Omega_{R} E_{\pm}
\label{eqn:chpt5a2}
\end{eqnarray}

Here $V=0.05$ is the strength of the inter-spin attraction relative to that of the like spin repulsion, $\gamma_e$ is the coherence decay rate and $\delta_e$ is the detuning of the excitonic resonance from the frequency of the pump field. $\Omega_R$ is the Rabi frequency, which couples the excitonic equations to the amplitudes $E_{\pm}$ of the $\sigma\pm$ circularly polarized components of the electric field. True photonic modes of the cavity have TE (subscript  $y$) and TM (subscript  $x$) symmetry. Amplitudes of these modes obey the following set of equations:

\begin{eqnarray}
&& \partial_tE_{x,y} -i \nabla^{2} E_{x,y} + (\gamma_{p} -i \delta_{p} ) E_{x,y} \\ &&
\nonumber  = i \Omega_R \psi_{x,y} + (a\pm b) E_p e^{i k_p x}.
\end{eqnarray}

While TE photonic and excitonic modes can be represented as $E_y= \frac{i}{\sqrt{2}}( E_+ - E_-)$, $\psi_y = \frac{i}{\sqrt{2}}( \psi_+ - \psi_- )$, the similar transformation for TM modes is only approximately valid, since the TM mode has a nonzero component perpendicular to the cavity plane, which does not couple to excitons. However, the external incident pump angle of $\sim20^{\circ}$, which is used here, corresponds to a small internal angle of $\sim6^{\circ}$ due to the high refractive index of GaAs. Therefore, the nonzero component of electric field perpendicular to the cavity plane is negligible, so the circularly polarised pump excites intracavity modes, which can be considered as circularly polarised.  Consequently, it is sufficient to assume that $E_x\simeq \frac{1}{\sqrt{2}}(E_+ +E_-)$, $\psi_x\simeq \frac{1}{\sqrt{2}}(\psi_+  + \psi_-$). In equation (2) $E_p$ and $k_p$ are the pump amplitude and transverse momentum while $\gamma_p$ and $\delta_p$ are the photon decay rate and detuning. Relative values of $a$ and $b$ ($|a|^2+|b|^2=1$) control the pump polarisation. %Small splittings of the TM and TE resonances, present for nonzero momenta, are irrelevant as shown below and are therefore disregarded in Eqs.~(1-2).

\begin{figure}
\centering
\includegraphics[width=8.5cm]{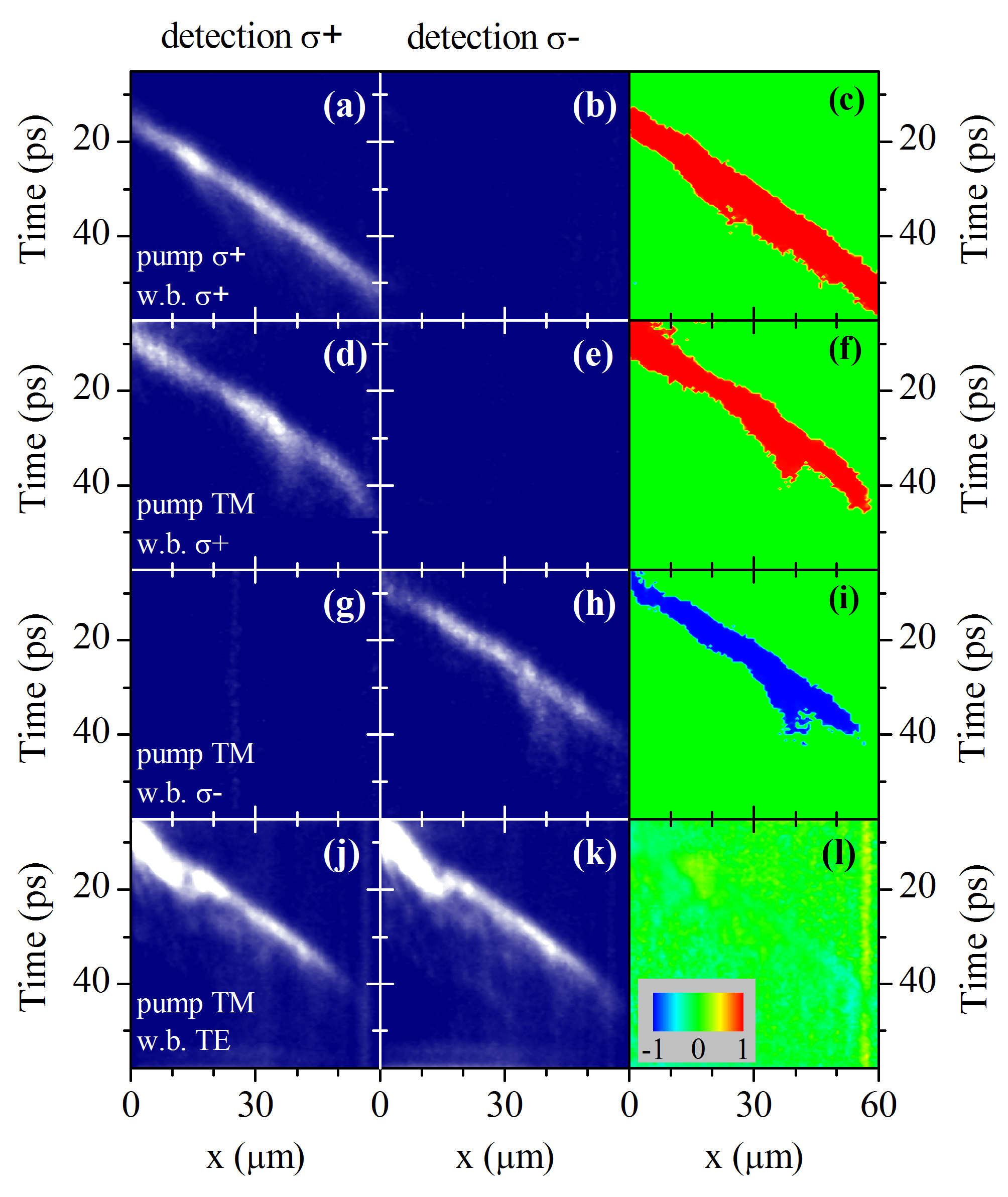}
\caption{(color online) Soliton emission intensity recorded as a function of time and position across the sample in $\sigma^+$ (a, d, g, j) and $\sigma^-$ (b, e, h, k) circular polarisations. Soliton CPD $\rho_c$ as function of time and position (c, f, i, l). The pump polarisation is $\sigma^+$ circular in (a-c) and TM linear in (d-l). The WB polarisation is $\sigma^+$ circular in (a-f), $\sigma^+$ circular in (g-i) and TE linear in (j-l).) The WB arrives at time t=0 ps at position $x\sim -10 $ $\upmu$m.
     }
\label{Fig1}
\end{figure}

\emph{Pump and WB circularly polarised. Experiment and theory ---} The bistability of pump polaritons \cite{tred,baas} is an important and well known prerequisite for the existence of solitons in microcavities \cite{amp}, since the latter can be considered as spatially localised excitations from a low intensity to a high intensity state of the bistability loop. Furthermore, the polariton-polariton stimulated parametric scattering process (modulation instability) from the switched on pump state also ensures that soliton harmonics with a broad range of k-vectors are populated.  Using our previous theoretical predictions \cite{prl11}, we tune the system into the bistability domain of its pump state and apply a co-circularly polarised WB, which excites a soliton. Experimentally measured spatio-temporal trajectories of such solitons and the temporal evolution of their  circular polarisation degree (CPD) $\rho_c$, defined as $\rho_c=(|E_+|^2-|E_-|^2)/(|E_+|^2+|E_-|^2)$, are shown in Figs.~1(a-c).  A high CPD of the soliton (Fig.~\ref{Fig1}(c)) is obtained with negligible emission in the opposing  circular polarisation component. Thus, when the pump and WB are circularly co-polarised, polariton-polariton scattering to the soliton harmonics occurs only in that polarisation

The full width half maximum (FWHM) of the soliton is measured to be $\sim$~6-7~$\upmu$m. Taking into account the resolution of our setup (5~$\upmu$m), we estimate a real soliton size of  $\sim $~3-4~$\upmu$m which is consistent with the healing length of the polariton fluid \cite{sich}.
We note that at low excitation densities there is a TE-TM splitting of polariton modes of $\sim$~0.1~meV at high k-vectors.
% which is neglected in the theory. 
This TE-TM splitting is responsible, for example, for the optical spin Hall effect \cite{spinHall} in the low density regime and for the formation of conservative dark spinor polariton solitons \cite{darkspin}. In a non-solitonic linear regime without a background CW pump, the polarisation of a polariton wavepacket excited at finite k-vectors would oscillate between $\sigma^+$ and $\sigma^-$ circular polarisations with a period of $\sim20$~ps. The fact that the $\rho_c$ is $\sim0.9$ over the whole duration of the soliton pulse $\sim35$~ps, as shown in Fig.~\ref{Fig1}(c), indicates effective quenching of the TE-TM splitting in the soliton regime. Phase synchronisation between the interacting TE and TM modes occurs since the soliton potential energy ($\sim$~0.3-0.5~meV) is greater than the TE-TM splitting \cite{sich}. Such a behaviour is closely related to recently observed quenching of the Zeeman splitting of polariton condensates arising from optical parametric oscillation \cite{Walker2011}. 
%The measurements of Fig.1 a)-c) demonstrate that despite the finite value of $k_p$, the microcavity field can be well described in the basis of circular polarisations.

In numerical modelling a circularly polarised pump is defined by setting $b=0$. If we take into account the TE-TM splitting of $\sim0.1$ meV  at k-vectors $k\sim k_p$ numerical simulations show excitation of circularly polarised solitons maintaining their polarisation during propagation in agreement with the experiment. For simplicity, in the theory analysis which follows next we disregarded this splitting.

%Thus, when the pump and WB are circularly co-polarised, polariton-polariton scattering to the soliton harmonics occurs only in that polarisation

\emph{Pump linearly polarised. Theory analysis of soliton stability---}Now we change the pump polarisation to linear, which corresponds to $b=a$ in Eqs.~(2), and tune our system into the bistability domain of the linearly polarised state. It was predicted and observed \cite{gipmult, Paraiso, WoutersMultiRes, Gavrilov2010, *Gavrilov2013} that if polaritons at nearly zero momentum are driven resonantly by a linearly polarised beam then different polarisation states of the spatially extended (non-solitonic) intracavity field ($\sigma+$, $\sigma+$ and linear) can all be stable in a finite range of pump powers. The situation is qualitatively different for polariton solitons. The thin lines in Fig.~2(a) show how the CPD of the internal homogeneous polariton field extending over the whole excitation pump spot (non-soliton regime) depends on $E_p$.  Linearly ($\rho_c=0$), elliptically and quasi-circularly polarised homogeneous polariton states may be excited for the same amplitude of the linearly polarised pump.  The $\rho_c$, $E_p$ relationship is symmetric with respect to $\rho_c\to -\rho_c$ due to the symmetry $(E_+,E_-)\to(E_-,E_+)$.  The quasi-circularly polarised spatially homogeneous states with $\rho_c$ close to $\pm 1$ are the only ones which are stable relative to perturbations with  momenta equal to the pump momentum. To find spatially localised soliton solutions, we transform Eqs.~(1, 2) into a reference frame moving with an unknown velocity $v$ and, by taking $\partial_t=\partial_y=0$, reduce them to a set of differential equations  with respect to the new coordinate $\xi=x-vt$. We then solve the resulting equations numerically and find the soliton profiles and associated velocities self-consistently \cite{sich}. Branches of stable (quasi-circularly polarised) and unstable (linearly and elliptically polarised) solitons are plotted in Fig.~2(a) as thick full and dashed lines, respectively. While repulsive interactions between co-circularly polarised polaritons keep the soliton population in resonance with the pump, the attractive coupling between $\sigma^+$ and $\sigma^-$ polarised polaritons pulls the system out of resonance.  This makes linearly polarised solitons unstable resulting in a break-up into circularly polarised components. Our numerical analysis shows that if these interactions are repulsive then linearly polarised solitons also become stable.

\begin{figure}
\includegraphics[width=8.5cm]{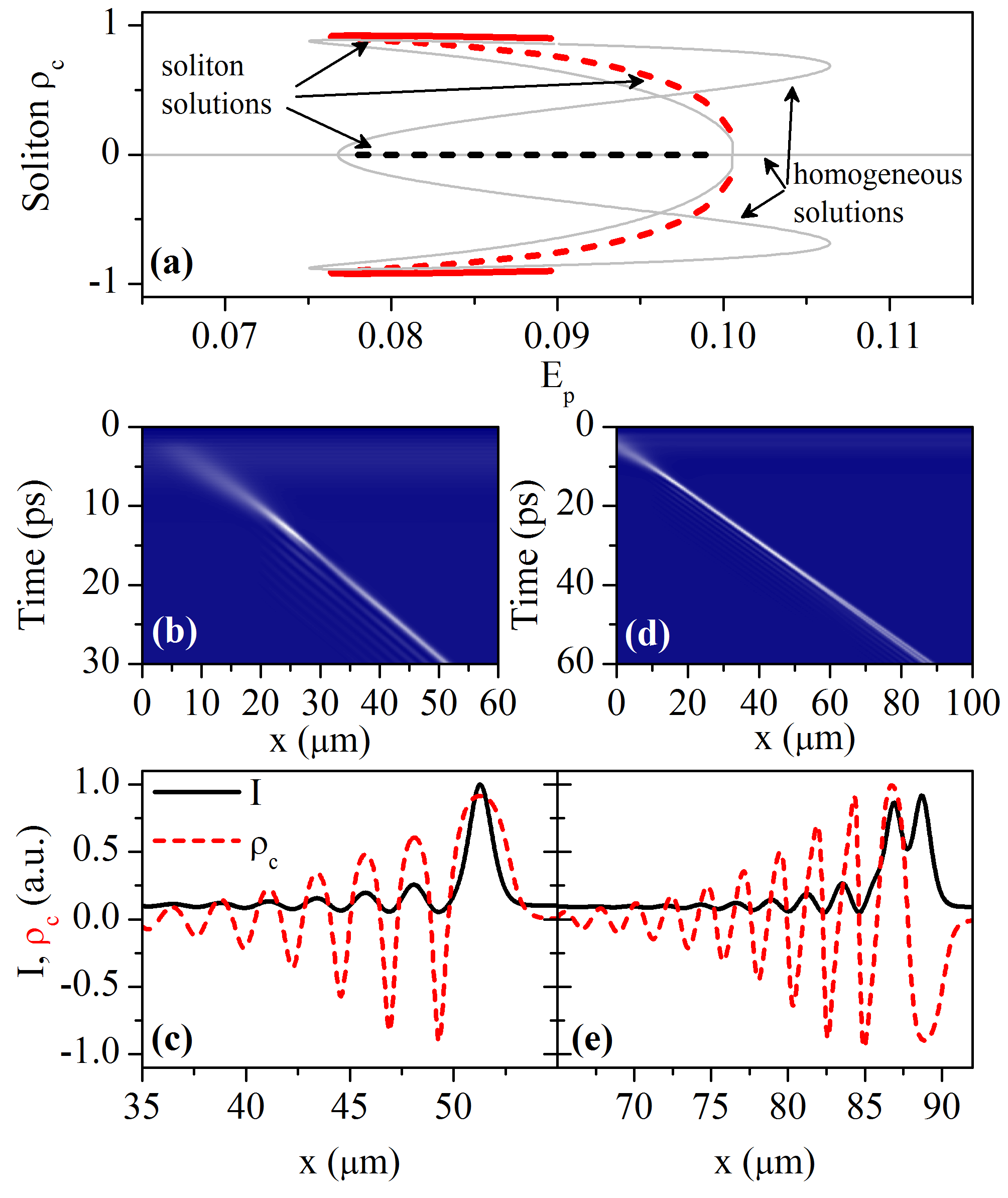}
\caption{(color online) (a) Thin grey lines show the CPD $\rho_c$ of polariton field vs pump amplitude corresponding to a homogeneous pump polariton field.  Dashed/solid lines correspond to unstable/stable soliton branches. (b), (d) Spatio-temporal trajectories of solitons when the pump and WB are orthogonally (b) and parallelly (d) linearly polarised. (c), (e) Soliton intensities and CPD $\rho_c$ as a function of position at time t=30 ps (c) and 60 ps (e) after the WB pulse. }
\label{Fig2}
\end{figure}

\emph{Pump linearly polarised, WB circularly polarised. Experiment---}
Stable solitons with $\sigma^+$ ($\sigma^-$) circular polarisation have been experimentally observed when a TM linearly polarised pump and a $\sigma^+$ ($\sigma^-$) circularly polarised WB are applied.  Figs.~1(d, e, g, h) show spatio-temporal traces of the soliton intensities measured in the $\sigma^+$ (d, g) and $\sigma^-$ (e, h) circular polarisations. We clearly observe soliton emission either in the $\sigma^+$ or $\sigma^-$ polarisations depending on the state of the WB. The measured $\rho_c$ for solitons is close to  $\pm 0.9$ in both cases as shown in  Fig.~1(f,i). When the $\sigma^+$ ($\sigma^-$) polarised WB arrives it causes blueshift of only the $\sigma^+$ ($\sigma^-$) polariton energies, bringing it into resonance with the pump. As a result, pure $\sigma^+$ ($\sigma^-$) polarised solitons are turned on. Importantly, such an effect is due to spin anisotropy in polariton-polariton interactions.  This is in agreement with our simulations, which show that if the spin-dependent polariton-polariton interactions were fully isotropic ($V=1$) then solitons co-polarised with the pump would be triggered independently of the WB polarisation. We note that we are able to generate either $\sigma^+$ or $\sigma^-$ polarised polariton solitons for the same linearly polarised pump depending on the WB, a possible basis for polarisation sensitive polariton based information processing schemes.

\emph{Pump and WB linearly polarised. Theory and experiment---}  When the polarisations of the pump and WB are orthogonal we find in the  simulations that instability of the linearly polarised solitons results in the development of  either $\sigma^+$ or $\sigma^-$ quasi-circularly polarised single solitons. These states can be excited with equal probabilities as the phases of the WB and pump are uncorrelated. An example of the calculated spatio-temporal trajectory of a quasi-circularly $\sigma^+$ polarised soliton is shown in Fig.~2(b). Fig.~2(c) shows the spatial distribution of the soliton intensity and CPD $\rho_c$ at time 30~ps. $\rho_c$ is close to 1 at the maximum of the soliton intensity, so that on average the soliton is circularly polarised. By contrast, in the case of quasi-parallel pump and WB polarisations, numerical modelling revealed the development of a double-hump soliton, where the polarisation changes from $\sigma^+$ to $\sigma^-$ between the intensity maxima on lengthscale of 2~$\upmu$m (see Figs.~2(d,e)).

Experimentally we were able to observe robust solitonic propagation only when the polarisations of the pump and WB are orthogonal \footnote{We were not able to observe solitons when the polarisations of the pump and WB were parallel, probably because of the increased WB and pump absorption due to excitation of biexcitons, which can only be excited with two co-linearly polarised photons}. The spatio-temporal trajectories of such solitons detected in $\sigma^+$ and $\sigma^-$ polarisations are shown in Fig.~1(j) and (k), respectively. The corresponding $\rho_c$ remains near zero as a function of time and position as shown in Fig.1 (l). Similar results are observed for measurements in the bases of horizontal(X)- vertical(Y) and diagonal(D1-D2) linear polarisations. The total polarisation degree $\rho=\sqrt{\rho_c^2+\rho_l^2}$ ($\rho_l$ is the linear polarisation degree) averaged over 10$^8$ pulses fluctuates near zero over the duration of the pulse. Such a result can be explained by either of the following two mechanisms: firstly,  the integrated CPD $\rho^{int}_c=\rho_c=( \int|E_+|^2dx-\int|E_-|^2dx )/(\int|E_+|^2dx+\int|E_-|^2dx)$ over the soliton size may be high, i.e a quasi-circularly polarised soliton is excited, but the polarisation orientation ($\sigma^+$ or $\sigma^-$) is random from pulse to pulse; secondly, both $\rho_c(x)$ as well as $\rho_l(x)$ may undergo rapid oscillations across the soliton profile similar to those in Fig.~2(e), so that on average $\rho^{int}_{c,l} \sim 0$. Such oscillations in the second case may also have random phase from pulse to pulse, leading to no observable spatial dependence of $\rho_{c,l}$ across the soliton profile averaged over many soliton pulses as in Fig.~1.

In order to distinguish between the two cases, we have recorded the number of $\sigma^+$ and $\sigma^-$ polarised photons emitted in each individual soliton pulse from a region of 5$\times$5 $\upmu$m$^2$ given by the resolution of our setup using a streak camera photon counting module \cite{corr}. We have computed the following correlation function of the soliton emission:
\begin{eqnarray}
g_c(m)=N\frac{\sum\limits_{i}{n^+(i)n^-(i+m)}}{\sum\limits_{i}{n^+(i)}\sum\limits_{i}{n^-(i)}},
\end{eqnarray}

Here $n^+(i)$ and $n^-(i+m)$ are the number of $\sigma^+$ and $\sigma^-$ polarised photons detected in the i-th and the (i+m)-th pulses, respectively. The averaging is performed over N=10$^5$  pulses. For a coherent soliton emission with a high integrated total polarisation degree $\rho^{int}=\sqrt{(\rho^{int}_c)^2+(\rho^{int}_l)^2}$, but totally random polarisation orientation from pulse to pulse, we expect $g_c(0)=0.66$ and $g_c(m>0)=1$. Similar results are expected for the correlation functions $g_c$ measured in X-Y  and D1-D2 bases. Surprisingly  all three correlation functions show values close to $1$ at $m=0$ within an error of 0.03 as shown in Fig.~3, implying that on average over the extent of the intracavity pulse the field is unpolarised. We associate such an observation with a rapid change of soliton polarisation in space on a scale of the order of the soliton size as in Fig.2 e). It is not clear
why such behaviour is observed for orthogonally polarised pump and w.b. in contrast to the theory, but we
speculate that it may arise from intrinsic photonic potential disorder across the sample.

\begin{figure}
\centering
\includegraphics[width=8.5cm]{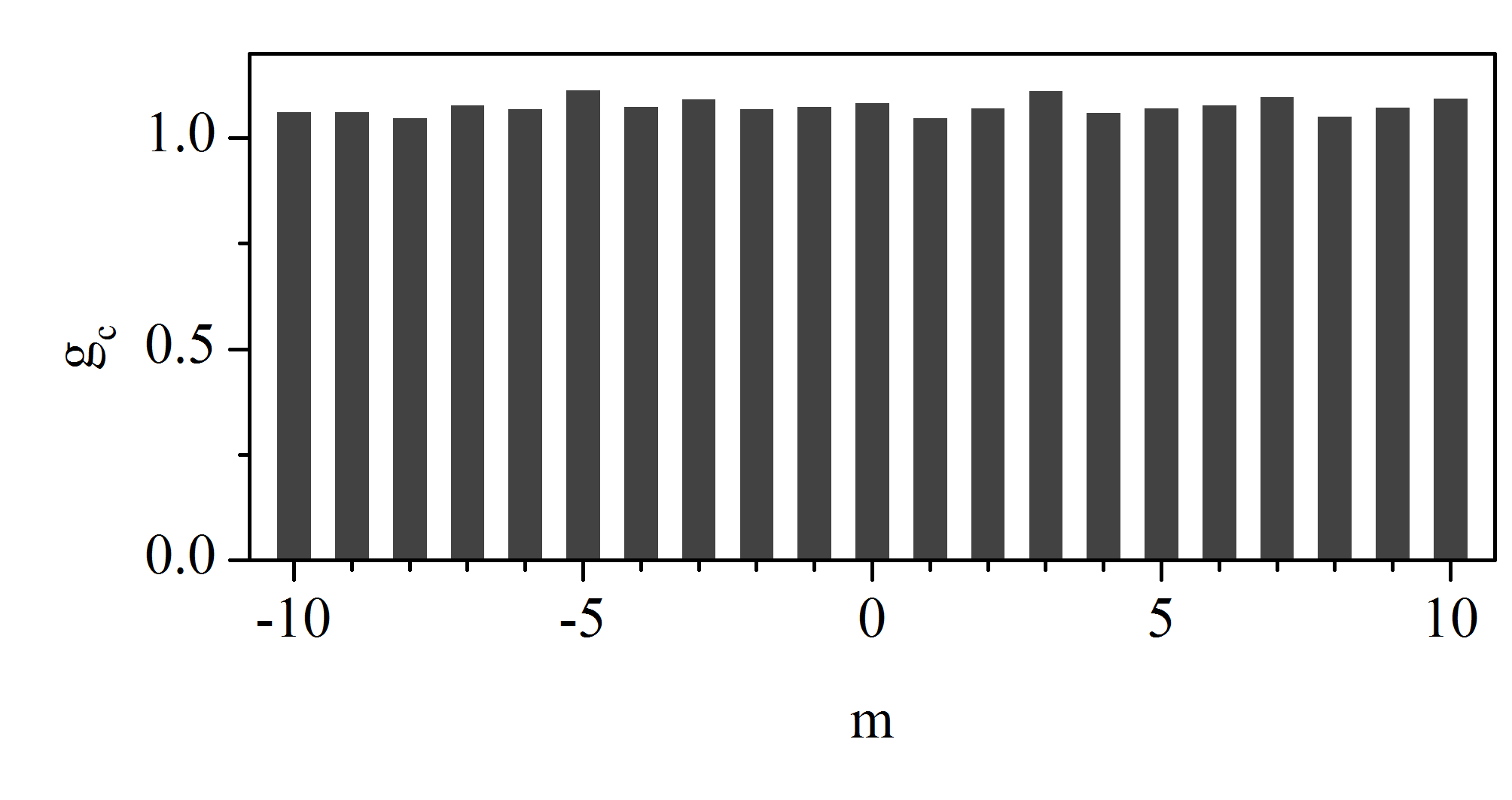}
\caption{ Correlations between cross-circularly polarised components of soliton emission. Correlation functions in the other bases X-Y and D1-D2 (not shown) are very similar. }
\label{Fig3}
\end{figure}

\emph{Dependence on CPD of pump. Experiment and theory---}We have also performed a systematic study of the soliton CPD $\rho_c$ as a function of the pump CPD $\rho_p$ for the cases of $\sigma^+$, $\sigma^-$ and linearly polarised WBs. The pump power was fixed at a maximum value of $\sim 120$~mW. The results, summarized in Fig.~4(a), show that the $\sigma^+$ ($\sigma^-$) polarised spatially localised pulses can be excited in the range of $\rho_p$ from 1 to -0.25 (-1 to 0.25) using $\sigma^+$ ($\sigma^-$) polarised WBs, so that there is a range $\rho_p$ from about -0.25 to 0.25 when both $\sigma^+$ and $\sigma^-$ polarised solitons can be triggered. For the  linearly polarised WB the slight bias of $\rho_p$ towards $\sigma^+$  or $\sigma^-$  polarisations by $0.1$ quickly ensures excitation of either $\sigma^+$  or $\sigma^-$  circularly polarised solitons with $|\rho| \sim 0.85$-0.95, implying the stability of circularly polarised solitons if the symmetry is slightly disturbed.  Numerically the exact stable solitons have been found by us only in a narrower intervals of $\rho_p$ at fixed pump power and pump angle (see Fig.~4(b)). In theory, in order to observe soliton emission at lower $\rho_p$ it is necessary to increase the pump field $E_p$ or the pump angle $\Theta_{pump}$. (Increasing $\Theta_{pump}$ brings CW pump close to polariton resonance and hence reduces the bistability threshold.) This is understandable since the pump bistability region at which the soliton forms is expected to shift to higher pump powers as $\rho_p$ decreases due to spin-dependent anisotropy in polariton-polariton interactions. Taking into account coupling of coherent excitons to the incoherent exciton reservoir through the formation and decay of bi-excitons \cite{Sarkar2010, WoutersMultiRes, Gavrilov2010a, *Gavrilov2012a} may lead to a better agreement between theory and experiment.

\begin{figure}
\centering
\includegraphics[width=8.5cm]{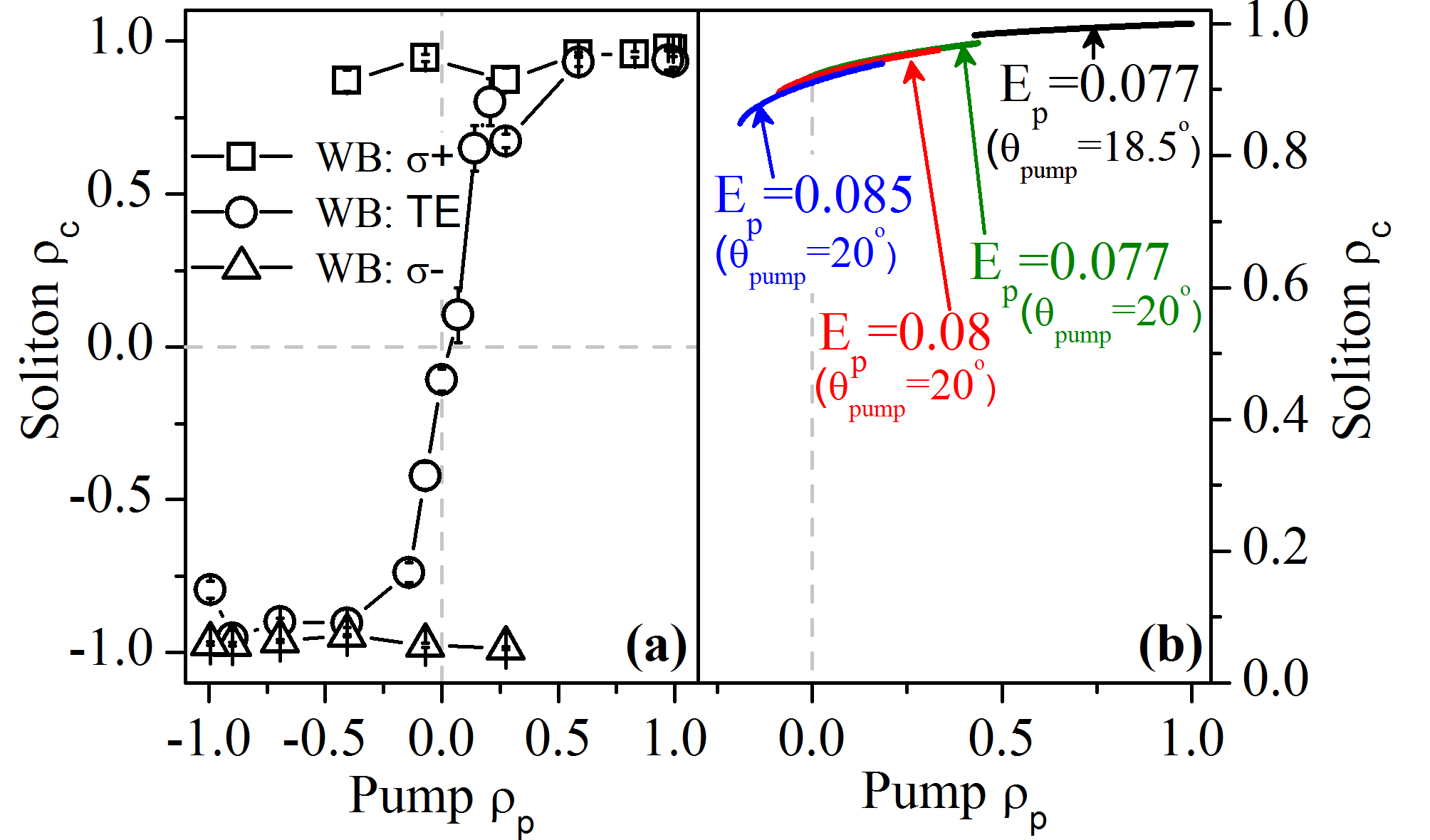}
\caption{ (a) Soliton CPD $\rho_c$ as a function of pump CPD $\rho_p$ recorded for the case of $\sigma^+$, $\sigma^-$ circularly polarised and TE linearly polarised WB (circles, squares and triangles, respectively). (b) numerical modelling of soliton CPD $\rho_c$ as a function of $\rho_p$ recorded for the case of $\sigma^+$ circularly polarised WB for different pump amplitudes and different pump angles.}
\label{Fig4}
\end{figure}

In conclusion, we have investigated the effect of spin-dependent polariton-polariton interactions on the spin properties of dissipative bright polariton solitons. Phase synchronisation between nondegenerate TE and TM polarised polariton modes ensures excitation of circularly polarised solitons which maintain their polarisation during the propagation. Stable $\sigma^+$ or $\sigma^-$ polarised solitons are observed under linearly polarised pumping, as the instability of linearly polarised solitons leads to fast evolution of the soliton polarisation in space-time as further supported by  correlation measurements. 
% Recently, a room temperature polariton laser has been demonstrated in GaN-based microcavities \cite{GaN}, while an electrically %pumped polariton laser has been realised in GaAs-based devices \cite{ellaser}.% These elements may make feasible the %construction of electrically pumped polariton soliton integrated circuits operating at 300 K.

We acknowledge support from EPSRC grant EP/J007544/1 and EU ITN grant "Clermont 4". D. Skryabin acknowledges support from Leverhulme Trust.


\begin{thebibliography}{45}%
\makeatletter
\providecommand \@ifxundefined [1]{%
 \@ifx{#1\undefined}
}%
\providecommand \@ifnum [1]{%
 \ifnum #1\expandafter \@firstoftwo
 \else \expandafter \@secondoftwo
 \fi
}%
\providecommand \@ifx [1]{%
 \ifx #1\expandafter \@firstoftwo
 \else \expandafter \@secondoftwo
 \fi
}%
\providecommand \natexlab [1]{#1}%
\providecommand \enquote  [1]{``#1''}%
\providecommand \bibnamefont  [1]{#1}%
\providecommand \bibfnamefont [1]{#1}%
\providecommand \citenamefont [1]{#1}%
\providecommand \href@noop [0]{\@secondoftwo}%
\providecommand \href [0]{\begingroup \@sanitize@url \@href}%
\providecommand \@href[1]{\@@startlink{#1}\@@href}%
\providecommand \@@href[1]{\endgroup#1\@@endlink}%
\providecommand \@sanitize@url [0]{\catcode `\\12\catcode `\$12\catcode
  `\&12\catcode `\#12\catcode `\^12\catcode `\_12\catcode `\%12\relax}%
\providecommand \@@startlink[1]{}%
\providecommand \@@endlink[0]{}%
\providecommand \url  [0]{\begingroup\@sanitize@url \@url }%
\providecommand \@url [1]{\endgroup\@href {#1}{\urlprefix }}%
\providecommand \urlprefix  [0]{URL }%
\providecommand \Eprint [0]{\href }%
\providecommand \doibase [0]{http://dx.doi.org/}%
\providecommand \selectlanguage [0]{\@gobble}%
\providecommand \bibinfo  [0]{\@secondoftwo}%
\providecommand \bibfield  [0]{\@secondoftwo}%
\providecommand \translation [1]{[#1]}%
\providecommand \BibitemOpen [0]{}%
\providecommand \bibitemStop [0]{}%
\providecommand \bibitemNoStop [0]{.\EOS\space}%
\providecommand \EOS [0]{\spacefactor3000\relax}%
\providecommand \BibitemShut  [1]{\csname bibitem#1\endcsname}%
\let\auto@bib@innerbib\@empty
%</preamble>
\bibitem [{\citenamefont {Kivshar}\ and\ \citenamefont {Agrawal}(2001)}]{agr}%
  \BibitemOpen
  \bibfield  {author} {\bibinfo {author} {\bibfnamefont {Y.}~\bibnamefont
  {Kivshar}}\ and\ \bibinfo {author} {\bibfnamefont {G.}~\bibnamefont
  {Agrawal}},\ }\href@noop {} {\emph {\bibinfo {title} {{Optical Solitons: From
  Fibers to Photonic Crystals}}}}\ (\bibinfo  {publisher} {Academic Press},\
  \bibinfo {year} {2001})\BibitemShut {NoStop}%
\bibitem [{\citenamefont {Ackemann}\ \emph {et~al.}(2009)\citenamefont
  {Ackemann} \emph {et~al.}}]{amp}%
  \BibitemOpen
  \bibfield  {author} {\bibinfo {author} {\bibfnamefont {T.}~\bibnamefont
  {Ackemann}} \emph {et~al.},\ }\href@noop {} {\emph {\bibinfo {title}
  {{Advances in Atomic, Molecular and Optical Physics}}}},\ Vol.~\bibinfo
  {volume} {57}\ (\bibinfo  {publisher} {Academic Press},\ \bibinfo {year}
  {2009})\ pp.\ \bibinfo {pages} {323--421}\BibitemShut {NoStop}%
\bibitem [{\citenamefont {Khaykovich}\ \emph {et~al.}(2002)\citenamefont
  {Khaykovich} \emph {et~al.}}]{becsol1}%
  \BibitemOpen
  \bibfield  {author} {\bibinfo {author} {\bibfnamefont {L.}~\bibnamefont
  {Khaykovich}} \emph {et~al.},\ }\href {\doibase 10.1126/science.1071021}
  {\bibfield  {journal} {\bibinfo  {journal} {Science}\ }\textbf {\bibinfo
  {volume} {296}},\ \bibinfo {pages} {1290} (\bibinfo {year}
  {2002})}\BibitemShut {NoStop}%
\bibitem [{\citenamefont {Strecker}\ \emph {et~al.}(2002)\citenamefont
  {Strecker} \emph {et~al.}}]{becsol2}%
  \BibitemOpen
  \bibfield  {author} {\bibinfo {author} {\bibfnamefont {K.~E.}\ \bibnamefont
  {Strecker}} \emph {et~al.},\ }\href {\doibase 10.1038/nature747} {\bibfield
  {journal} {\bibinfo  {journal} {Nature}\ }\textbf {\bibinfo {volume} {417}},\
  \bibinfo {pages} {150} (\bibinfo {year} {2002})}\BibitemShut {NoStop}%
\bibitem [{\citenamefont {Kavokin}\ \emph {et~al.}(2007)\citenamefont {Kavokin}
  \emph {et~al.}}]{book}%
  \BibitemOpen
  \bibfield  {author} {\bibinfo {author} {\bibfnamefont {A.}~\bibnamefont
  {Kavokin}} \emph {et~al.},\ }\href@noop {} {\emph {\bibinfo {title}
  {{Microcavities}}}},\ Semiconductor science and technology\ (\bibinfo
  {publisher} {Oxford University Press},\ \bibinfo {year} {2007})\BibitemShut
  {NoStop}%
\bibitem [{\citenamefont {Sanvitto}\ and\ \citenamefont
  {Timofeev}(2012)}]{book2}%
  \BibitemOpen
  \bibfield  {author} {\bibinfo {author} {\bibfnamefont {D.}~\bibnamefont
  {Sanvitto}}\ and\ \bibinfo {author} {\bibfnamefont {V.~B.}\ \bibnamefont
  {Timofeev}},\ }\href@noop {} {\emph {\bibinfo {title} {{Exciton Polaritons in
  Microcavities: New Frontiers}}}},\ \bibinfo {series} {Solid State Sciences},
  Vol.\ \bibinfo {volume} {172}\ (\bibinfo  {publisher} {Springer},\ \bibinfo
  {year} {2012})\BibitemShut {NoStop}%
\bibitem [{\citenamefont {Krizhanovskii}\ \emph {et~al.}(2008)\citenamefont
  {Krizhanovskii} \emph {et~al.}}]{krizh2000}%
  \BibitemOpen
  \bibfield  {author} {\bibinfo {author} {\bibfnamefont {D.}~\bibnamefont
  {Krizhanovskii}} \emph {et~al.},\ }\href {\doibase
  10.1103/PhysRevB.77.115336} {\bibfield  {journal} {\bibinfo  {journal} {Phys.
  Rev. B}\ }\textbf {\bibinfo {volume} {77}},\ \bibinfo {pages} {115336}
  (\bibinfo {year} {2008})}\BibitemShut {NoStop}%
\bibitem [{\citenamefont {Gippius}\ \emph {et~al.}(2004)\citenamefont {Gippius}
  \emph {et~al.}}]{gip}%
  \BibitemOpen
  \bibfield  {author} {\bibinfo {author} {\bibfnamefont {N.~A.}\ \bibnamefont
  {Gippius}} \emph {et~al.},\ }\href {\doibase 10.1209/epl/i2004-10133-6}
  {\bibfield  {journal} {\bibinfo  {journal} {Europhys. Lett.}\ }\textbf
  {\bibinfo {volume} {67}},\ \bibinfo {pages} {997} (\bibinfo {year}
  {2004})}\BibitemShut {NoStop}%
\bibitem [{\citenamefont {Tredicucci}\ \emph {et~al.}(1996)\citenamefont
  {Tredicucci} \emph {et~al.}}]{tred}%
  \BibitemOpen
  \bibfield  {author} {\bibinfo {author} {\bibfnamefont {A.}~\bibnamefont
  {Tredicucci}} \emph {et~al.},\ }\href@noop {} {\bibfield  {journal} {\bibinfo
   {journal} {Phys.l Rev. A}\ }\textbf {\bibinfo {volume} {54}},\ \bibinfo
  {pages} {3493} (\bibinfo {year} {1996})}\BibitemShut {NoStop}%
\bibitem [{\citenamefont {Baas}\ \emph {et~al.}(2004)\citenamefont {Baas} \emph
  {et~al.}}]{baas}%
  \BibitemOpen
  \bibfield  {author} {\bibinfo {author} {\bibfnamefont {A.}~\bibnamefont
  {Baas}} \emph {et~al.},\ }\href {\doibase 10.1103/PhysRevA.69.023809}
  {\bibfield  {journal} {\bibinfo  {journal} {Phys. Rev. A}\ }\textbf {\bibinfo
  {volume} {69}},\ \bibinfo {pages} {023809} (\bibinfo {year}
  {2004})}\BibitemShut {NoStop}%
\bibitem [{\citenamefont {Kasprzak}\ \emph {et~al.}(2006)\citenamefont
  {Kasprzak} \emph {et~al.}}]{kasprzak}%
  \BibitemOpen
  \bibfield  {author} {\bibinfo {author} {\bibfnamefont {J.}~\bibnamefont
  {Kasprzak}} \emph {et~al.},\ }\href {\doibase 10.1038/nature05131} {\bibfield
   {journal} {\bibinfo  {journal} {Nature}\ }\textbf {\bibinfo {volume}
  {443}},\ \bibinfo {pages} {409} (\bibinfo {year} {2006})}\BibitemShut
  {NoStop}%
\bibitem [{\citenamefont {Balili}\ \emph {et~al.}(2007)\citenamefont {Balili}
  \emph {et~al.}}]{balili}%
  \BibitemOpen
  \bibfield  {author} {\bibinfo {author} {\bibfnamefont {R.}~\bibnamefont
  {Balili}} \emph {et~al.},\ }\href {\doibase 10.1126/science.1140990}
  {\bibfield  {journal} {\bibinfo  {journal} {Science}\ }\textbf {\bibinfo
  {volume} {316}},\ \bibinfo {pages} {1007} (\bibinfo {year}
  {2007})}\BibitemShut {NoStop}%
\bibitem [{\citenamefont {Amo}\ \emph {et~al.}(2009)\citenamefont {Amo} \emph
  {et~al.}}]{AmoNP2009}%
  \BibitemOpen
  \bibfield  {author} {\bibinfo {author} {\bibfnamefont {A.}~\bibnamefont
  {Amo}} \emph {et~al.},\ }\href {\doibase 10.1038/NPHYS1364} {\bibfield
  {journal} {\bibinfo  {journal} {Nature Phys.}\ }\textbf {\bibinfo {volume}
  {5}},\ \bibinfo {pages} {805} (\bibinfo {year} {2009})}\BibitemShut {NoStop}%
\bibitem [{\citenamefont {Yulin}\ \emph {et~al.}(2008)\citenamefont {Yulin}
  \emph {et~al.}}]{yulin}%
  \BibitemOpen
  \bibfield  {author} {\bibinfo {author} {\bibfnamefont {A.}~\bibnamefont
  {Yulin}} \emph {et~al.},\ }\href@noop {} {\bibfield  {journal} {\bibinfo
  {journal} {Phys. Rev. A}\ }\textbf {\bibinfo {volume} {78}},\ \bibinfo
  {pages} {061801} (\bibinfo {year} {2008})}\BibitemShut {NoStop}%
\bibitem [{\citenamefont {Amo}\ \emph {et~al.}(2011)\citenamefont {Amo} \emph
  {et~al.}}]{amo}%
  \BibitemOpen
  \bibfield  {author} {\bibinfo {author} {\bibfnamefont {A.}~\bibnamefont
  {Amo}} \emph {et~al.},\ }\href {\doibase 10.1126/science.1202307} {\bibfield
  {journal} {\bibinfo  {journal} {Science}\ }\textbf {\bibinfo {volume}
  {332}},\ \bibinfo {pages} {1167} (\bibinfo {year} {2011})}\BibitemShut
  {NoStop}%
\bibitem [{\citenamefont {Adrados}\ \emph {et~al.}(2011)\citenamefont {Adrados}
  \emph {et~al.}}]{Adrados2011}%
  \BibitemOpen
  \bibfield  {author} {\bibinfo {author} {\bibfnamefont {C.}~\bibnamefont
  {Adrados}} \emph {et~al.},\ }\href {\doibase 10.1103/PhysRevLett.107.146402}
  {\bibfield  {journal} {\bibinfo  {journal} {Phys. Rev. Lett.}\ }\textbf
  {\bibinfo {volume} {107}},\ \bibinfo {pages} {146402} (\bibinfo {year}
  {2011})}\BibitemShut {NoStop}%
\bibitem [{\citenamefont {Egorov}\ \emph {et~al.}(2009)\citenamefont {Egorov}
  \emph {et~al.}}]{prl11}%
  \BibitemOpen
  \bibfield  {author} {\bibinfo {author} {\bibfnamefont {O.}~\bibnamefont
  {Egorov}} \emph {et~al.},\ }\href {\doibase 10.1103/PhysRevLett.102.153904}
  {\bibfield  {journal} {\bibinfo  {journal} {Phys. Rev. Lett.}\ }\textbf
  {\bibinfo {volume} {102}},\ \bibinfo {pages} {153904} (\bibinfo {year}
  {2009})}\BibitemShut {NoStop}%
\bibitem [{\citenamefont {Gorbach}\ \emph {et~al.}(2009)\citenamefont {Gorbach}
  \emph {et~al.}}]{pla}%
  \BibitemOpen
  \bibfield  {author} {\bibinfo {author} {\bibfnamefont {A.}~\bibnamefont
  {Gorbach}} \emph {et~al.},\ }\href@noop {} {\bibfield  {journal} {\bibinfo
  {journal} {Phys. Lett. A}\ }\textbf {\bibinfo {volume} {373}},\ \bibinfo
  {pages} {3024} (\bibinfo {year} {2009})}\BibitemShut {NoStop}%
\bibitem [{\citenamefont {Sich}\ \emph {et~al.}(2012)\citenamefont {Sich} \emph
  {et~al.}}]{sich}%
  \BibitemOpen
  \bibfield  {author} {\bibinfo {author} {\bibfnamefont {M.}~\bibnamefont
  {Sich}} \emph {et~al.},\ }\href {\doibase 10.1038/NPHOTON.2011.267}
  {\bibfield  {journal} {\bibinfo  {journal} {Nature Photon.}\ }\textbf
  {\bibinfo {volume} {6}},\ \bibinfo {pages} {50} (\bibinfo {year}
  {2012})}\BibitemShut {NoStop}%
\bibitem [{\citenamefont {Tanese}\ \emph {et~al.}(2013)\citenamefont {Tanese}
  \emph {et~al.}}]{blochncomm}%
  \BibitemOpen
  \bibfield  {author} {\bibinfo {author} {\bibfnamefont {D.}~\bibnamefont
  {Tanese}} \emph {et~al.},\ }\href {\doibase 10.1038/ncomms2760} {\bibfield
  {journal} {\bibinfo  {journal} {Nature Comm.}\ }\textbf {\bibinfo {volume}
  {4}},\ \bibinfo {pages} {1749} (\bibinfo {year} {2013})}\BibitemShut
  {NoStop}%
\bibitem [{\citenamefont {Boyd}(2003)}]{boyd}%
  \BibitemOpen
  \bibfield  {author} {\bibinfo {author} {\bibfnamefont {R.~W.}\ \bibnamefont
  {Boyd}},\ }\href@noop {} {\emph {\bibinfo {title} {{Nonlinear Optics}}}}\
  (\bibinfo  {publisher} {Academic Press},\ \bibinfo {year} {2003})\BibitemShut
  {NoStop}%
\bibitem [{\citenamefont {Barland}\ \emph {et~al.}(2002)\citenamefont {Barland}
  \emph {et~al.}}]{barland}%
  \BibitemOpen
  \bibfield  {author} {\bibinfo {author} {\bibfnamefont {S.}~\bibnamefont
  {Barland}} \emph {et~al.},\ }\href {\doibase 10.1038/nature01114.1.}
  {\bibfield  {journal} {\bibinfo  {journal} {Nature}\ }\textbf {\bibinfo
  {volume} {419}},\ \bibinfo {pages} {699} (\bibinfo {year}
  {2002})}\BibitemShut {NoStop}%
\bibitem [{\citenamefont {Pedaci}\ \emph {et~al.}(2008)\citenamefont {Pedaci}
  \emph {et~al.}}]{pedaci}%
  \BibitemOpen
  \bibfield  {author} {\bibinfo {author} {\bibfnamefont {F.}~\bibnamefont
  {Pedaci}} \emph {et~al.},\ }\href {\doibase 10.1063/1.2828458} {\bibfield
  {journal} {\bibinfo  {journal} {Appl. Phys. Lett.}\ }\textbf {\bibinfo
  {volume} {92}},\ \bibinfo {pages} {011101} (\bibinfo {year}
  {2008})}\BibitemShut {NoStop}%
\bibitem [{\citenamefont {Zutic}\ \emph {et~al.}(2004)\citenamefont {Zutic}
  \emph {et~al.}}]{Sarma}%
  \BibitemOpen
  \bibfield  {author} {\bibinfo {author} {\bibfnamefont {I.}~\bibnamefont
  {Zutic}} \emph {et~al.},\ }\href@noop {} {\bibfield  {journal} {\bibinfo
  {journal} {Rev. Mod. Phys.}\ }\textbf {\bibinfo {volume} {76}},\ \bibinfo
  {pages} {323} (\bibinfo {year} {2004})}\BibitemShut {NoStop}%
\bibitem [{\citenamefont {Liew}\ \emph {et~al.}(2008)\citenamefont {Liew} \emph
  {et~al.}}]{LiewNeurons}%
  \BibitemOpen
  \bibfield  {author} {\bibinfo {author} {\bibfnamefont {T.}~\bibnamefont
  {Liew}} \emph {et~al.},\ }\href {\doibase 10.1103/PhysRevLett.101.016402}
  {\bibfield  {journal} {\bibinfo  {journal} {Phys. Rev. Lett.}\ }\textbf
  {\bibinfo {volume} {101}},\ \bibinfo {pages} {016402} (\bibinfo {year}
  {2008})}\BibitemShut {NoStop}%
\bibitem [{\citenamefont {Gippius}\ \emph {et~al.}(2007)\citenamefont {Gippius}
  \emph {et~al.}}]{gipmult}%
  \BibitemOpen
  \bibfield  {author} {\bibinfo {author} {\bibfnamefont {N.}~\bibnamefont
  {Gippius}} \emph {et~al.},\ }\href {\doibase 10.1103/PhysRevLett.98.236401}
  {\bibfield  {journal} {\bibinfo  {journal} {Phys. Rev. Lett.}\ }\textbf
  {\bibinfo {volume} {98}},\ \bibinfo {pages} {236401} (\bibinfo {year}
  {2007})}\BibitemShut {NoStop}%
\bibitem [{\citenamefont {Sarkar}\ \emph {et~al.}(2010)\citenamefont {Sarkar}
  \emph {et~al.}}]{Sarkar2010}%
  \BibitemOpen
  \bibfield  {author} {\bibinfo {author} {\bibfnamefont {D.}~\bibnamefont
  {Sarkar}} \emph {et~al.},\ }\href {\doibase 10.1103/PhysRevLett.105.216402}
  {\bibfield  {journal} {\bibinfo  {journal} {Phys. Rev. Lett.}\ }\textbf
  {\bibinfo {volume} {105}},\ \bibinfo {pages} {216402} (\bibinfo {year}
  {2010})}\BibitemShut {NoStop}%
\bibitem [{\citenamefont {Para\"{\i}so}\ \emph {et~al.}(2010)\citenamefont
  {Para\"{\i}so} \emph {et~al.}}]{Paraiso}%
  \BibitemOpen
  \bibfield  {author} {\bibinfo {author} {\bibfnamefont {T.~K.}\ \bibnamefont
  {Para\"{\i}so}} \emph {et~al.},\ }\href@noop {} {\bibfield  {journal}
  {\bibinfo  {journal} {Nature Mat.}\ }\textbf {\bibinfo {volume} {9}},\
  \bibinfo {pages} {655} (\bibinfo {year} {2010})}\BibitemShut {NoStop}%
\bibitem [{\citenamefont {Kitano}\ \emph {et~al.}(1981)\citenamefont {Kitano}
  \emph {et~al.}}]{kitano}%
  \BibitemOpen
  \bibfield  {author} {\bibinfo {author} {\bibfnamefont {M.}~\bibnamefont
  {Kitano}} \emph {et~al.},\ }\href@noop {} {\bibfield  {journal} {\bibinfo
  {journal} {Phys. Rev. Lett.}\ }\textbf {\bibinfo {volume} {46}},\ \bibinfo
  {pages} {926} (\bibinfo {year} {1981})}\BibitemShut {NoStop}%
\bibitem [{\citenamefont {Cecchi}\ \emph {et~al.}(1982)\citenamefont {Cecchi}
  \emph {et~al.}}]{cecchi}%
  \BibitemOpen
  \bibfield  {author} {\bibinfo {author} {\bibfnamefont {S.}~\bibnamefont
  {Cecchi}} \emph {et~al.},\ }\href@noop {} {\bibfield  {journal} {\bibinfo
  {journal} {Phys. Rev. Lett.}\ }\textbf {\bibinfo {volume} {49}},\ \bibinfo
  {pages} {1928} (\bibinfo {year} {1982})}\BibitemShut {NoStop}%
\bibitem [{\citenamefont {Shelykh}\ \emph {et~al.}(2008)\citenamefont {Shelykh}
  \emph {et~al.}}]{shel}%
  \BibitemOpen
  \bibfield  {author} {\bibinfo {author} {\bibfnamefont {I.}~\bibnamefont
  {Shelykh}} \emph {et~al.},\ }\href {\doibase 10.1103/PhysRevLett.100.116401}
  {\bibfield  {journal} {\bibinfo  {journal} {Phys. Rev. Lett.}\ }\textbf
  {\bibinfo {volume} {100}},\ \bibinfo {pages} {116401} (\bibinfo {year}
  {2008})}\BibitemShut {NoStop}%
\bibitem [{\citenamefont {Amo}\ \emph {et~al.}(2010)\citenamefont {Amo} \emph
  {et~al.}}]{AmoNPhot2010}%
  \BibitemOpen
  \bibfield  {author} {\bibinfo {author} {\bibfnamefont {A.}~\bibnamefont
  {Amo}} \emph {et~al.},\ }\href {\doibase 10.1038/nphoton.2010.79} {\bibfield
  {journal} {\bibinfo  {journal} {Nature Photon.}\ }\textbf {\bibinfo {volume}
  {4}},\ \bibinfo {pages} {361} (\bibinfo {year} {2010})}\BibitemShut {NoStop}%
\bibitem [{\citenamefont {Gavrilov}\ \emph
  {et~al.}(2013{\natexlab{a}})\citenamefont {Gavrilov} \emph
  {et~al.}}]{Gavrilov2013a}%
  \BibitemOpen
  \bibfield  {author} {\bibinfo {author} {\bibfnamefont {S.~S.}\ \bibnamefont
  {Gavrilov}} \emph {et~al.},\ }\href {\doibase 10.1103/PhysRevB.87.201303}
  {\bibfield  {journal} {\bibinfo  {journal} {Phys. Rev. B}\ }\textbf {\bibinfo
  {volume} {87}},\ \bibinfo {pages} {201303} (\bibinfo {year}
  {2013}{\natexlab{a}})}\BibitemShut {NoStop}%
\bibitem [{\citenamefont {Hivet}\ \emph {et~al.}(2012)\citenamefont {Hivet}
  \emph {et~al.}}]{darkspin}%
  \BibitemOpen
  \bibfield  {author} {\bibinfo {author} {\bibfnamefont {R.}~\bibnamefont
  {Hivet}} \emph {et~al.},\ }\href {\doibase 10.1038/nphys2406} {\bibfield
  {journal} {\bibinfo  {journal} {Nature Phys.}\ }\textbf {\bibinfo {volume}
  {8}},\ \bibinfo {pages} {724} (\bibinfo {year} {2012})}\BibitemShut {NoStop}%
\bibitem [{\citenamefont {Leyder}\ \emph {et~al.}(2007)\citenamefont {Leyder}
  \emph {et~al.}}]{spinHall}%
  \BibitemOpen
  \bibfield  {author} {\bibinfo {author} {\bibfnamefont {C.}~\bibnamefont
  {Leyder}} \emph {et~al.},\ }\href {\doibase doi:10.1038/nphys676} {\bibfield
  {journal} {\bibinfo  {journal} {Nature Phys.}\ }\textbf {\bibinfo {volume}
  {3}},\ \bibinfo {pages} {628} (\bibinfo {year} {2007})}\BibitemShut {NoStop}%
\bibitem [{\citenamefont {Walker}\ \emph {et~al.}(2011)\citenamefont {Walker}
  \emph {et~al.}}]{Walker2011}%
  \BibitemOpen
  \bibfield  {author} {\bibinfo {author} {\bibfnamefont {P.}~\bibnamefont
  {Walker}} \emph {et~al.},\ }\href {\doibase 10.1103/PhysRevLett.106.257401}
  {\bibfield  {journal} {\bibinfo  {journal} {Phys. Rev. Lett.}\ }\textbf
  {\bibinfo {volume} {106}},\ \bibinfo {pages} {257401} (\bibinfo {year}
  {2011})}\BibitemShut {NoStop}%
\bibitem [{\citenamefont {Wouters}\ \emph {et~al.}(2013)\citenamefont {Wouters}
  \emph {et~al.}}]{WoutersMultiRes}%
  \BibitemOpen
  \bibfield  {author} {\bibinfo {author} {\bibfnamefont {M.}~\bibnamefont
  {Wouters}} \emph {et~al.},\ }\href {\doibase 10.1103/PhysRevB.87.045303}
  {\bibfield  {journal} {\bibinfo  {journal} {Phys. Rev. B}\ }\textbf {\bibinfo
  {volume} {87}},\ \bibinfo {pages} {045303} (\bibinfo {year}
  {2013})}\BibitemShut {NoStop}%
\bibitem [{\citenamefont {Gavrilov}\ \emph
  {et~al.}(2010{\natexlab{a}})\citenamefont {Gavrilov} \emph
  {et~al.}}]{Gavrilov2010}%
  \BibitemOpen
  \bibfield  {author} {\bibinfo {author} {\bibfnamefont {S.}~\bibnamefont
  {Gavrilov}} \emph {et~al.},\ }\href {\doibase 10.1134/S1063776110050146}
  {\bibfield  {journal} {\bibinfo  {journal} {J. Exp. Theor. Phys}\ }\textbf
  {\bibinfo {volume} {110}},\ \bibinfo {pages} {825} (\bibinfo {year}
  {2010}{\natexlab{a}})}\BibitemShut {NoStop}%
\bibitem [{\citenamefont {Gavrilov}\ \emph
  {et~al.}(2013{\natexlab{b}})\citenamefont {Gavrilov} \emph
  {et~al.}}]{Gavrilov2013}%
  \BibitemOpen
  \bibfield  {author} {\bibinfo {author} {\bibfnamefont {S.~S.}\ \bibnamefont
  {Gavrilov}} \emph {et~al.},\ }\href {\doibase 10.1063/1.4773523} {\bibfield
  {journal} {\bibinfo  {journal} {Appl. Phys. Lett.}\ }\textbf {\bibinfo
  {volume} {102}},\ \bibinfo {pages} {011104} (\bibinfo {year}
  {2013}{\natexlab{b}})}\BibitemShut {NoStop}%
\bibitem [{Note1()}]{Note1}%
  \BibitemOpen
  \bibinfo {note} {We were not able to observe solitons when the polarisations
  of the pump and WB were parallel, probably because of the increased WB and
  pump absorption due to excitation of biexcitons, which can only be excited
  with two co-linearly polarised photons}\BibitemShut {NoStop}%
\bibitem [{\citenamefont {Wiersig}\ \emph {et~al.}(2009)\citenamefont {Wiersig}
  \emph {et~al.}}]{corr}%
  \BibitemOpen
  \bibfield  {author} {\bibinfo {author} {\bibfnamefont {J.}~\bibnamefont
  {Wiersig}} \emph {et~al.},\ }\href {\doibase 10.1038/nature08126} {\bibfield
  {journal} {\bibinfo  {journal} {Nature}\ }\textbf {\bibinfo {volume} {460}},\
  \bibinfo {pages} {245} (\bibinfo {year} {2009})}\BibitemShut {NoStop}%
\bibitem [{\citenamefont {Gavrilov}\ \emph
  {et~al.}(2010{\natexlab{b}})\citenamefont {Gavrilov} \emph
  {et~al.}}]{Gavrilov2010a}%
  \BibitemOpen
  \bibfield  {author} {\bibinfo {author} {\bibfnamefont {S.~S.}\ \bibnamefont
  {Gavrilov}} \emph {et~al.},\ }\href {\doibase 10.1134/S0021364010150105}
  {\bibfield  {journal} {\bibinfo  {journal} {J. Exp. Theor. Phys Lett.}\
  }\textbf {\bibinfo {volume} {92}},\ \bibinfo {pages} {171} (\bibinfo {year}
  {2010}{\natexlab{b}})}\BibitemShut {NoStop}%
\bibitem [{\citenamefont {Gavrilov}\ \emph {et~al.}(2012)\citenamefont
  {Gavrilov} \emph {et~al.}}]{Gavrilov2012a}%
  \BibitemOpen
  \bibfield  {author} {\bibinfo {author} {\bibfnamefont {S.~S.}\ \bibnamefont
  {Gavrilov}} \emph {et~al.},\ }\href {\doibase 10.1103/PhysRevB.85.075319}
  {\bibfield  {journal} {\bibinfo  {journal} {Phys. Rev. B}\ }\textbf {\bibinfo
  {volume} {85}},\ \bibinfo {pages} {075319} (\bibinfo {year}
  {2012})}\BibitemShut {NoStop}%
%\bibitem [{\citenamefont {Christmann}\ \emph {et~al.}(2008)\citenamefont
  %{Christmann} \emph {et~al.}}]{GaN}%
  %\BibitemOpen
  %\bibfield  {author} {\bibinfo {author} {\bibfnamefont {G.}~\bibnamefont
  %{Christmann}} \emph {et~al.},\ }\href {\doibase 10.1063/1.2966369} {\bibfield
   %{journal} {\bibinfo  {journal} {Appl. Phys. Lett.}\ }\textbf {\bibinfo
  %{volume} {93}},\ \bibinfo {pages} {051102} (\bibinfo {year}
  %{2008})}\BibitemShut {NoStop}%
%\bibitem [{\citenamefont {Schneider}\ \emph {et~al.}(2013)\citenamefont
  %{Schneider} \emph {et~al.}}]{ellaser}%
  %\BibitemOpen
 % \bibfield  {author} {\bibinfo {author} {\bibfnamefont {C.}~\bibnamefont
  %{Schneider}} \emph {et~al.},\ }\href {\doibase 10.1038/nature12036}
  %{\bibfield  {journal} {\bibinfo  {journal} {Nature}\ }\textbf {\bibinfo
  %{volume} {497}},\ \bibinfo {pages} {348} (\bibinfo {year}
  %{2013})}\BibitemShut {NoStop}%
\end{thebibliography}
\end{document}